\documentclass[aps,pra,twocolumn,showpacs,superscriptaddress]{revtex4-1}

\usepackage{amsfonts,mathrsfs,amsmath,amsthm,amssymb}
\usepackage{bm,times}
\usepackage{graphicx,epsfig}
\usepackage[colorlinks=true,breaklinks=true,linkcolor=blue,citecolor=blue,urlcolor=blue]{hyperref}

\hypersetup{
  pdfauthor={Tao Zhou},
  pdfkeywords={},
  pdftitle={},
  pdfsubject={Research Paper},
  pdfpagemode=UseNone
}

\def \non {\nonumber}

\def \ket {\rangle}
\def \bra {\langle}

\newcommand{\be}{\begin{eqnarray}}
\newcommand{\ee}{\end{eqnarray}}

\newcommand{\im}{\mathrm{i}}

\makeatletter
    
    \newcommand{\Rmnum}[1]{\expandafter\@slowromancap\romannumeral #1@}
\makeatother

\begin{document}

\title{Pryce's mass-center operators and the anomalous velocity of a spinning electron}
\author{Long Huang}
\author{Xiaohua Wu}
\email{wxhscu@scu.edu.cn}
\affiliation{College of Physical Science and Technology, Sichuan University, Chengdu 610064, China}
\author{Tao Zhou}
\email{taozhou@swjtu.edu.cn}
\affiliation{School of Physical Science and Technology, Southwest Jiaotong University, Chengdu 610031, China}

\date{\today}

\begin{abstract}
In the present work, we develop a method to calculate the anomalous velocity of a spinning electron. From Dirac equation, the relationships among the expectation values of the Pryce's mass-center operator, the position operator, the spin operator and the canonical momentum operator are investigated.  By requiring that the center of mass for the classical spinning electron is related to the expectation value of the Pryce's mass-center operator, one can obtain a classical expression for the position of the electron. With the classical equations of motion, the anomalous velocity of a spinning electron can be easily calculated. It is shown that two factors contribute to the anomalous velocity: one is dependent on the selection of the Pryce's mass-center operators and the other is a type-independent velocity expressed by the rotational velocity and the Lorentz force.
\end{abstract}

\pacs{03.65.Ca, 03.30.+p}

\maketitle

\section{introduction}
\label{intro}

The equations of motion for Bloch electrons in electromagnetic fields were constructed by Bloch, Peierls, Jones and Zener around 1930~\cite{Bloch,Peierls1,Jones101}, and have played a fundamental role in the physics of metals and semiconductors. In the tight-band
model, Peierls~\cite{Peierls2} showed that the effective Hamiltonian in the presence of a magnetic field can be obtained by using the gauge invariant momentum operator in the unperturbed band energy, instead of the crystal momentum. Later, a more rigorous derivation of the effective Hamiltonian was given by Slater~\cite{PhysRev.76.1592} and Luttinger~\cite{PhysRev.84.814}, and was extended to many-body operator formalism by Adams~\cite{PhysRev.85.41}. A correction term to the velocity, known as the anomalous velocity, was introduced by Karplus, Luttinger, and Kohn~\cite{PhysRev.95.1154,*PhysRev.108.590,*PhysRev.112.739}. Chang and Niu~\cite{PhysRevLett.75.1348} connected the anomalous velocity correction to the Berry phase with the electron motion in an energy band. The semiclassical theory of one-band in the wave-packet formulation was constructed in Ref.~\cite{PhysRevB.53.7010,*PhysRevB.59.14915}, and the multi-band wave packet formulation was also developed in Ref.~\cite{PhysRevB.72.085110,SHINDOU2005399}. For the details of this topic, see Ref.~\cite{RevModPhys.82.1959} and references therein.

Besides the known methods to construct the nonrelativistic quantum Hamiltonian, Chang and Niu proposed a different way to derive the effective Hamiltonian. With the gauge-invariant semiclassical theory, the effective Hamiltonian can be quantized using the generalized Peierls substitution~\cite{0953-8984-20-19-193202}. To get a better grasp of the theoretical formulation, the authors have taken the relativistic particle as an illustrative example. For a Dirac wave packet constructed from free-particle states with positive energy, it has been shown that in
the presence of the electromagnetic field, the mass center of the wave packet has a non-vanishing anomalous velocity.

Recently, in a semiclassical kinetic theory where Dirac fermion is studied within the matrix differential form method, Dayi and Kilincarslan~\cite{DAYI2015119} showed that the kinematic Thomas precession correction should be taken into account. The Thomas precession contributes to the one-form obtained by the semiclassical wave packet on an equal footing with the Berry gauge field, and it yields the cancellation of the anomalous velocity terms when ignoring the higher terms in momentum. The relation between spin and the Berry-phase contribution to the anomalous velocity of massive and massless Dirac particle has also be considered by Stone, Dwivedi, and Zhou~\cite{PhysRevD.91.025004}, where they introduced a covariant Berry connection and investigated how it enters the classical relativistic dynamics of spinning particles. For the massive Dirac particle with $g=2$, they concluded that the anomalous velocity correction does not exist.

A quantum description of spin is based on the Dirac equation, while the most popular classical equations of the spinning electron were formulated by Frenkel~\cite{Fenkel} and Bargmann, Michel and Telegdi (BMT)~\cite{PhysRevLett.2.435}. In the work by Fradkin and Good~\cite{RevModPhys.33.343}, it was shown how the electron's classical equation of motion, the Lorentz force equation and the BMT-equation, are related to the Dirac equation. As we show in the following, the Fradkin-Good-protocol (FG-protocol) involves both the definition of the spin operators for Dirac wave packet and the derivation of the relations among the expectation values of different operators, and by introducing a set of substitution rules, the known classical equations are recovered. 

In the present work, we suggest that the anomalous velocity can be studied in a different protocol. Our basic assumption is that the mass center of a classical spinning electron should be associated with the expectation value of a mass-center operator for the Dirac wave packet. Unfortunately, contrary to the non-relativistic cases, where the mass center can be easily defined, it is a highly nontrivial task to define the mass center of many-particles system in relativity. Therefore, Pryce~\cite{Pryce62} proposed three possible mass-center position operators labeled by (c),(d) and (e) and relevant spin operators can be constructed. In 1949, Newton and Wigner~\cite{RevModPhys.21.400} found the Pryce's e-type operator in the investigation of localized states for elementary systems, and they also showed that Pryce's e-type operator is the only position operator with commuting components in the Dirac theory which has localized eigenfunctions in the manifold of positive energy wave functions. Later on, the Newton-Wigner position operator, or the Pryce's e-type mass-center operator, is called the mean-position operator by Foldy and Wouthuysen~\cite{PhysRev.78.29}, and it is showed that when the particle interacts with an external field, it is the mean-position operator that is identified with the position operator in the non-relativistic Pauli theory.

As an application of the FG-protocol, we derive a relation among the expectation values of the mass-center operator, the spin operator, the position operator and the canonical momentum operator. With the substitution rule, one can obtain a classical expression for the mass center of
a spinning electron. From the known classical equations of motion, it is shown that the contributions to the anomalous velocity of the electron comes from two aspects: a type-dependent factor and a type-independent velocity expressed by the rotational velocity and the Lorentz force. 

The content of present work is organized as follows. In Sec.~\ref{sec2}, a brief review of the FG-protocol is introduced. In Sec.~\ref{sec3}, a relation among the expectation values of different operators are derived for the Pryce's mass-center operators. In Sec.~\ref{sec4}, with the introduced substitution rule, we obtain a model for the mass ceter of a spinning electron. In Sec.~\ref{sec5}, applying the classic equations of motion, we derive the anomalous velocity of the electron. Finally, we end our work with a short discussion in Sec.~\ref{sec6}.

\section{The Fradkin-Good protocol}
\label{sec2}

For a point-like particle, its world line is denoted by $(t, \bm x)$. In this paper, we set $\hbar=c=1$. The $4$-momentum $(E,\bm p)$ for a massive particle is
\be
E=\bar{\gamma} m,\ \ \bm p=\bar{\gamma} m \bm v,\non
\ee
where $\bar{\gamma}=1/\sqrt{(1-{\bm v}^2)}$ is the dilation factor, and $\bm v=d\bm x/dt$.  The Dirac matrices $\gamma_\mu$ ($\mu=1,2,3,4$) are defined by
\be
\gamma_{\mu}\gamma_{\nu}+\gamma_{\nu}\gamma_{\mu}=\delta_{\mu\nu},\non
\ee
and the auxiliary matrices $\gamma_5$, $\beta$, $\alpha_i$, $\sigma_i$ ($i=1,2,3$) can be introduced as 
\be
\gamma_5&=&\gamma_1\gamma_2\gamma_3\gamma_4,\ \ \beta=\gamma_4,\nonumber\\
\alpha_i&=&\im\beta\gamma_i,\ \ \ \sigma_i=\im\gamma_4\gamma_5\gamma_i.\nonumber
\ee
For convenience, a specific representation is chosen as
\be
\bm\alpha=\left(
               \begin{array}{cc}
                 0 & \bm\sigma \\
                  \bm\sigma & 0\\
               \end{array}
             \right),\beta=\left(
                             \begin{array}{cc}
                               1 & 0 \\
                               0 & -1\\
                             \end{array}
                           \right),\gamma_5=\left(
                                              \begin{array}{cc}
                                                0& -1\\
                                                -1 & 0 \\
                                              \end{array}
                                            \right),\nonumber
\ee
with the $2\times 2$ Pauli matrices $\bm\sigma$,
\be
\sigma_1=\left(
           \begin{array}{cc}
             0 & 1 \\
             1 & 0 \\
           \end{array}
         \right),
\sigma_2=\left(
 \begin{array}{cc}
0 & -\im \\
\im& 0 \\
\end{array}
\right), 
\sigma_3=\left(
\begin{array}{cc}
1 & 0 \\
0 & -1 \\
\end{array}
\right).\nonumber
\ee
[In this paper, $\bm\sigma$ may be a $4\times4$ or a $2\times2$ matrix, which is determinated according to the actual cases].

The free Dirac equation for spin-$1/2$ particle is
\be
\im\frac{\partial \Psi}{\partial t}=\hat{H}\Psi,\non
\ee
with the Hamitonian $\hat{H}=\bm\alpha\cdot\hat{\bm p}+\beta m$. In the classic work~\cite{PhysRev.78.29}, Foldy and Wouthuysen introduced a unitary transformation,
\be
e^{\pm\im\hat{S}} =\frac{E\pm\beta\bm\alpha\cdot \bm p+m}{[2E(E+m)]^{\frac{1}{2}}},\non
\ee
where $\hat{S}$ is a Hermitian operator and $E=\sqrt{m^2+\bm p^2}$. It is showed that the Newton-Wigner position operator $\hat{\bm X}_{\mathrm{NW}}$,
which is also known as the Pryce's e-type operator or the mean-position operator, is related to the position operator $\hat{\bm x}$ in the transformed representation,
\be
\hat{\bm X}_{\mathrm{NW}}=e^{-\im\hat{S}}\hat{\bm x}e^{\im\hat{S}}.\non
\ee
A similar result is obtained by Fradkin and Good~\cite{RevModPhys.33.343}, where they attempted to give a consistent account of electron  polarization, and a 3-vector  operator $\bm O$ in the FW-representation is defined
\be
\bm O=e^{-\im\hat{S}}\beta\bm\sigma e^{\im\hat{S}}.\non
\ee
The operator $\bm O$ has a deep relation with the operators $T_\mu$ ($\mu=1,2,3,4$)
\be
\bm T=\beta\bm\sigma-\gamma_5\hat{\bm p},\ \ {T}_4 =\im\bm \sigma\cdot\hat{\bm p},\non
\ee
with $\hat{\bm p}=-\im\bm \nabla$, and $T_\mu$ are the generators of the little group: a subgroup of homogeneous Lorentz transformations that leaves the  4-vector momentum of a plane-wave state unchanged. [In the original work~\cite{RevModPhys.33.343}, the mass $m$ is set to be unit, and in the following, we recovere it to be a non-unit parameter.] Before one can show how these operators are related, we should first give a brief review of $\mathrm{FG}$'s protocol to get the relation among the expectation values of different operators.

For a Hermitian operator $Q$, its expectation value is defined as
\be
\bra Q\ket=\int d^3\bm x\Psi^\dag Q\Psi,\non
\ee
where the integration extends overall space. In electrodynamics, the field is described by the vector-potential $(\phi, \bm A)$, and $\bm E$ and $\bm B$ are the electric and magnetic filed, respectively. When the electron is interacted with the electromagnetic field, the $3$-momentum operator $\hat{\bm p}$ should be replaced by the canonical momentum operator $\hat{\bm\pi}$,
\be
\label{sub}
\hat{\bm p}\rightarrow\hat{\bm\pi}=\hat{\bm p}-e\bm A,
\ee
with $e$ the charge of the electron. Therefore, one can have
\be
E=\sqrt{\hat{\bm\pi}^2+m^2},\ \ \bm T=\beta\bm\sigma-\gamma_5{\hat{\bm\pi}},\ \ T_4 =\im\bm\sigma\cdot\hat{\bm\pi},\non
\ee
and the free Hamilton operator $\hat{H}$ is generalized to
\be
\hat{H}=\bm\alpha\cdot\hat{\bm\pi}+\beta m +e\phi.\non
\ee

In the case where the wave function $\Psi(\bm x,t)$ is negligible except in a small region, the following rules (a), (b) and (c) are quite usefull in FG-protocol:

(a) The wave function is supposed to have a sharp spread in momentum and energy value, and then,
\be
\hat{\pi}_{\mu}\Psi(\bm x,t)=\bra\hat{\pi}_\mu\rangle\Psi(\bm x,t),\non
\ee
where $\bra\hat{\pi}_\mu\ket$ is the classical value varying along the orbit but factorable out of the integration on the wave function. [Note $\hat{\pi}_0=-\im\hat{\pi}_4=\sqrt{\bra\hat{\pi}\ket^2+m^2}$].

(b) $e\bm B/m$ and $e\bm E/m$ are negligible to unit, and then, the expectation value of $\hat{H}-e\phi$ is
\be
\bra \hat{H}-e\phi\ket=E(\bra\hat{\bm\pi}\ket),\non\\
E(\bra\hat{\bm\pi}\ket)=\bar{\gamma}m, \bm v=\frac{\bra\hat{\bm\pi}\ket}{\bar{\gamma} m}.\non
\ee

(c) For any Hermitian operator $Q$,
\be
\bra[Q,\hat{H}-e\phi]_+\ket=2\bar{\gamma}m\bra Q\ket,\non
\ee
with $[A,B]_+=AB+BA$, and this equation is useful to determining the classical equations of motion.

In the following, the $\mathrm{FG}$'s protocol is appllied to derive the classical equation of the position operator $\hat{\bm x}$. The operator equation, $d\hat{\bm x}/dt=\im[\hat{H}, \hat{\bm x}]$ is useful, and with the results $[\hat{H},\hat{\bm x}]=-\im\bm \alpha$ and $\bra\bm \alpha\ket=\bm v$, one can have
\be
\frac{d\bra \hat{\bm x}\ket}{dt}=\bm v.\non
\ee
Similarly, one can also obtain
\be
\label{eq2}
\frac{d\bra\hat{\bm \pi}\ket}{dt}=e(\bm E+\bm v\times\bm B).
\ee

For the operators $T_\mu$ and $\bm O$, employing the substitution rule in Eq.~(\ref{sub}), Fradkin and Good acquired a series of relations
\be
\bra\bm T\ket&=&\bra\bm O\ket+\frac{\bar{\gamma}^2}{\bar{\gamma}+1}(\bm v\cdot\bra\bm O\ket)\bm v,\\
\bra T_4\ket&=&\im\bar{\gamma}\bm v\cdot\bra\bm O\ket,\\
\bra\bm O\ket&=&\bra\bm T\ket-\frac{\bar{\gamma}}{\bar{\gamma}+1}(\bm v\cdot\bra\bm T\ket)\bm v,\\
\frac{d\bra\bm O\ket}{dt}&=&\frac{e}{\bar{\gamma}m}\bra\bm O\ket\times\bigg[\bm B+\frac{\bar{\gamma}}{\bar{\gamma}+1}\bm E\times\bm v\bigg].
\label{eq3}
\ee

\section{Pryce's mass-center operators}
\label{sec3}

According to the substitution rule in Eq.~(\ref{sub}), the Pryce's mass-center operators take the forms~\cite{Pryce62}:
\be
\hat{\bm X}_{\mathrm{P}}^{(\mathrm d)}&=&\hat{\bm x}+\frac{\im\beta\bm\alpha}{2m}-\frac{\im\beta(\bm\alpha\cdot\hat{\bm\pi})\hat{\bm\pi}}{2mE^2},\non\\
\hat{\bm X}_{\mathrm{P}}^\mathrm{(e)}&=&\hat{\bm x}+\frac{\im\beta\bm \alpha}{2E}+\frac{\hat{\bm\pi}\times\bm\sigma}{2E(E+m)}-\frac{\im\beta(\bm \alpha\cdot\hat{\bm\pi})\hat{\bm\pi}}{2E^2(E+m)},\non\\
\hat{\bm X}_{\mathrm{P}}^\mathrm{(c)}&=&\hat{\bm x}+\frac{\im m\beta\bm\alpha}{2E^2}+\frac{\hat{\bm\pi}\times\bm\sigma}{2E^2}.\non
\ee
With the protocol introduced above, the expectation values of the Pryce's  mass-center operators  can be expressed in terms of $\bra\hat{\bm x}\ket$, $\bra\bm T\ket$ and $\bra\hat{\bm\pi}\ket$. Some results in the following have already been obtained by Fradkin and Good, and we would like to give a simple proof for them. First, with the two equations $[\im\beta\bm\alpha,\hat{H}-e\phi]_{+}=2 (\beta\bm\sigma)\times\hat{\bm\pi}$,  $\bra \beta\bm\sigma\ket\times\bra\hat{\bm\pi}\ket=\bra\bm T\ket\times\bra\hat{\bm\pi}\ket$ and rule (c), one can come to
\be
\label{eq1}
\bra\im\beta\bm\alpha\ket=(\bar{\gamma} m)^{-1}\bra\bm T\ket\times\bra\hat{\bm\pi}\ket,
\ee
and this is the Eq. (17.11) in FG's work~\cite{RevModPhys.33.343}. In the same way, with $[\bm\sigma, \hat{H}-e\phi]_+=2m\bm T$, one can obtain
\be
\bra\bm\sigma\ket=\bra\bm T\ket/\bar{\gamma},\non
\ee
and this is the Eq. (17.8) in the original work~\cite{RevModPhys.33.343}. According to rule (a), $\bra\hat{\bm\pi}\times\bm\sigma\ket=\bra\hat{\bm\pi}\ket\times\bra\bm\sigma\ket$, and therefore,
\be
\bra\hat{\bm\pi}\times\bm\sigma\ket=\bra\hat{\bm\pi}\ket\times\bra\bm T\ket/\bar{\gamma}.\non
\ee
Finally, use Eq.~(\ref{eq1}), follow the argument in rule (a), and one has
\be
\bra\im\beta(\bm\alpha\cdot\hat{\bm\pi})\hat{\bm\pi}\ket=0.\non
\ee

Based on results above, it is convenient for us to define a general mass-center operator,
\be
\hat{\bm X}_{\mathrm{P}}^{(i)}=\hat{\bm x}+f^{(i)}_1\frac{\im\beta\bm\alpha}{2m}+f^{(i)}_2\frac{\hat{\bm\pi}\times\bm\sigma}{2m^2}+f^{(i)}_3\frac{\im\beta(\bm\alpha\cdot\hat{\bm\pi})\hat{\bm\pi}}{2m^3}\non
\ee
with $i=\mathrm{d, e, c}$. The type-dependent factors $f^{(i)}_j$ are known to be
\be
f_1^\mathrm{(d)}&=&1, f_2^\mathrm{(d)}=0, f_3^\mathrm{(d)}=-\frac{1}{\bar{\gamma}^2},\non\\
f_1^\mathrm{(e)}&=&\frac{1}{\bar{\gamma}}, f_2^\mathrm{(e)}=\frac{1}{\bar{\gamma}(1+\bar{\gamma})},   f_3^\mathrm{(e)}=-\frac{1}{\bar{{\gamma}}^2(\bar{{\gamma}}+1)},\non\\
f_1^\mathrm{(c)}&=&\frac{1}{\bar{\gamma}^2}, f_2^\mathrm{(c)}=\frac{1}{\bar{\gamma}^2}, f_3^\mathrm{(c)}=0.\non
\ee
Together with the results above, one of the main result in present work appears,
\be
\bra\hat{\bm X}_{\mathrm{P}}^{(i)}\ket=\bra\hat{\bm x}\ket+f_{\mathrm{P}}^{(i)}\frac{\bra\bm T\ket\times\bra\hat{\bm\pi}\ket}{2m^2\bar{\gamma}},\non
\ee
where $f_{\mathrm{P}}^{(i)}$ are type-dependent factors defined by
\be
f_{\mathrm{P}}^{(i)}=f_1^{(i)}-f_2^{(i)}.\non
\ee
In a more explicit form, they should be
\be
\label{eq8}
f_{\mathrm{P}}^\mathrm{(d)}=1,f_{\mathrm{P}}^\mathrm{(e)}=\frac{1}{1+\bar{\gamma}},f_{\mathrm{P}}^\mathrm{(c)}=0.
\ee

\section{Classic equations of motion}
\label{sec4}

Beside the Lorentz force equation for the electron~\cite{Jackson}
\be
\label{eq4}
\frac{d\bm p}{dt}=e[\bm E+\bm v\times\bm B],
\ee
the polarization of electron in the rest frame, denoted by a 3-vector $\bm s$, is described by the $\mathrm{BMT}$ equation
\be
\frac{d\bm s}{dt}=\bm s\times\bm\omega,
\ee
where $\bm\omega$ is the rotational velocity of the polarization $\bm s$,
\be
\bm\omega=\frac{e}{m\bar{\gamma}}\bigg[\bm B+\frac{\bar{\gamma}}{1+\bar{\gamma}}\bm E\times\bm v\bigg].
\ee
In an inertial frame where the particle's velocity is $\bm v$, a $4$-vector spin $(S_0, \bm S)$ can be introduced
\be
S_0=\bar{\gamma}\bm v\cdot\bm s,\ \ \bm S=\bm s+\frac{\bar{\gamma}^2}{\bar{\gamma}+1}(\bm s\cdot\bm v)\bm v,
\ee
and therefore,
\be
\label{eq5}
\bm s=\bm S-\frac{\bar{\gamma}}{\bar{\gamma}+1}(\bm v\cdot \bm S)\bm v.
\ee
Obviously, by the substitution rule, one can have
\be
\label{eq6}
\bra\hat{\bm x}\ket=\bm x, \bra T_4\ket=\im S_0, \bra\hat{\bm\pi}\ket=\bm p, \bra\bm T\ket=\bm S, \bra\bm O\ket=\bm s,
\ee
the results from Eq.~(\ref{eq2}) to Eq.~(\ref{eq3}) are equivalent to the ones from Eq.~(\ref{eq4}) to Eq.~(\ref{eq5}). In the present work, besides the substitution rule in Eq.~(\ref{eq6}), it is required that the position (mass center) of a classical spinning electron is determined once one of the Pryce mass-center operators is selected, say
\be
\label{eq7}
\bm X_{\mathrm{P}}^{(i)}=\bra\hat{\bm X}_{\mathrm{P}}^{(i)}\ket.
\ee
For convenience, one can introduce a type-independent $3$-vector, the so-called position shift $\delta\bm X_{\mathrm{P}}$,
\be
\label{eq10}
\delta\bm X_{\mathrm{P}}=\frac{1}{2m}\bm S\times\bm v,
\ee
and Eq.~(\ref{eq7}) can be expressed into an explicit form,
\be
\bm X_{\mathrm{P}}^{(i)}=\bm x+f^{(i)}_{\mathrm{P}}(\bar{\gamma})\delta \bm X_{\mathrm{P}},\non
\ee
where the type-dependent factors $f^{(i)}_{\mathrm{P}}(\bar{\gamma})$ are given in Eq.~(\ref{eq8}).

\section{anomalous velocity}
\label{sec5}

In this section, we suppose that the energy of the spinning electron, $E=\bar{\gamma} m$, is changing slowly enough with time. In other words, it is required that
\be
\frac{d\bm v^2}{dt}=0,\non
\ee
and obviously, there should be $d\bar{\gamma}/dt=0$. Since the type-dependent factors $f^{(i)}_{\mathrm{P}}(\bar{\gamma})$ depend only on the dilation factor $\bar{\gamma}$, we may just focus on the position shift for simplicity, and a type-independent anomalous velocity $\bm V_{\mathrm{P}}$ can be defined as
\be
\bm V_{\mathrm{P}}=\frac{d}{dt}\delta\bm X_{\mathrm{P}},\non
\ee
The velocity of the mass center  $\bm X_{\mathrm{P}}^{(i)}$ can be known if the anomalous velocity $\bm V_{\mathrm{P}}$ has been determined, say
\be
\frac{d}{dt}\bm X_{\mathrm{P}}^{(i)}=\bm v+f^{(i)}_{\mathrm{P}}\bm V_{\mathrm{P}}.\non
\ee

Take advantage of the following intermediate equations,
\be
\frac{d\bm S}{dt}\times\bm v&=&\frac{d\bm s}{dt}\times\bm v+(\bm s\cdot\bm v)\bm\omega_{\mathrm{T}},\non\\
\bm S\times\frac{d\bm v}{dt}&=&\bm s\times\frac{d\bm v}{dt}-(\bm s\cdot\bm v)\bm \omega_{\mathrm{T}},\nonumber
\ee
where $\bm \omega_{\mathrm{T}}$ is the Thomas precession~\cite{Thomas},
\be
\bm \omega_{\mathrm{T}}=\frac{\bar{\gamma}^2}{\bar{\gamma}+1}\frac{d\bm v}{dt}\times\bm v,\non
\ee
one can come to a more compact expression of the anomalous velocity $\bm V_{\mathrm{P}}$,
\be
\label{eq9}
\bm V_{\mathrm{P}}=\frac{1}{2m}\frac{d(\bm s\times\bm v)}{dt}.
\ee
Based on the classic equations of motion and with the identity equation $\bm a\times (\bm b\times\bm c)=(\bm a\cdot\bm c)\bm b-(\bm a\cdot \bm b)\bm c$, the anomalous velocity  can be further expressed as
\be
\bm V_{\mathrm{P}}=\frac{1}{2m}[(\bm s\cdot\bm v)\bm\omega-(\bm\omega\cdot\bm v)\bm s+\bm s\times\frac{\bm F}{m}],\non
\ee
with $\bm F=m d\bm v/dt$ the Lorentz force.

Following the results in Ref.~\cite{Jackson}, the evolution of the $3$-vector $\bm s$ can be also described by
\be
\frac{d\bm s}{dt}=\frac{1}{\bar{\gamma}}\bm F'+\bm \omega_{\mathrm{T}}\times\bm s,\non
\ee
where
\be
\bm F'=\frac{\bar{\gamma}ge}{2m}\bm s\times\bigg[\bm B-\frac{\bar{\gamma}}{1+\bar{\gamma}}(\bm v\cdot\bm B)\bm v-\bm v\times\bm E\bigg].\non
\ee
[In this paper, the $g$ factor is  $g=2$].  Therefore, under the condition that $\bm F'=0$,
\be
\label{eq11}
\bm V_{\mathrm{P}}=\frac{1}{2m}\bigg[-(\bm s\cdot\bm v)\bm\omega_\mathrm{T}+\bm s\times\frac{\bm F}{m}\bigg],
\ee
and it is clear that the anomalous velocity is related to both the Thomas precession and the Lorentz force.

Furthermore, the anomalous velocity can be decomposed into two terms:
\be
\bm V_{\mathrm{P}}=\bm V_{\mathrm{P}}(\bm E)+\bm V_{\mathrm{P}}(\bm B),\non
\ee
where 
\be
\bm V_{\mathrm{P}}(\bm E)&=&\frac{e}{2m^2\bar{\gamma}}\bigg[\bm s-\frac{\bar{\gamma}}{1+\bar{\gamma}}(\bm s\cdot\bm v)\bm v\bigg]\times \bm E,\non\\
\bm V_{\mathrm{P}}(\bm B)&=&\frac{e}{2m^2\bar{\gamma}}\bigg[(\bm s\cdot\bm B)\bm v-(\bm v\cdot\bm B)\bm s\bigg].\non
\ee
can be derived from the definition in Eq.~(\ref{eq9}) and the classical equations of motion.

In the low velocity limit, $\bar{\gamma}\approx1$, one can keep terms to the first order of $\bm v$, and
\be
\bm V_{\mathrm{P}}(\bm E)=\frac{e}{2m^2}\bm s\times\bm E.\non
\ee
When the magnetic filed is absent, $\bm B=0$, the results yield
\be
\frac{d}{dt}\bm X_{\mathrm{P}}^\mathrm{(d)}&=&\bm v+\frac{e}{2m^2}\bm s\times\bm E,\non\\
\frac{d}{dt}\bm X_{\mathrm{P}}^\mathrm{(e)}&=&\bm v+\frac{e}{4m^2}\bm s\times\bm E,\non\\
\frac{d}{dt}\bm X_{\mathrm{P}}^\mathrm{(c)}&=&\bm v.\non
\ee
Obviously, the actual value of the anomalous velocity is dependent on the choice of the Pryce's mass-center operators.

\section{Remarks and discussion}
\label{sec6}

As we show in this work, the anomalous velocity of a spinning electron is determined by the type-independent velocity $\bm V_{\mathrm{P}}$ and the choice of the Pryce's mass-center operators. In the low velocity limit, the anomalous velocity for the Pryce's e-type operator is half of the one for the d-type operator, while the anomalous velocity for the c-type mass-center operator is always zero.

When one tries to construct a classical model for the mass center of a relativistic spinning electron, he may encounter an elementary problem in defining the mass-center operator for the Dirac wave packet, and the same situation also happened in other works. One example is in the construction of the Lagrangian for Frenkel electron~\cite{Deriglazov2014}. Another example appears in the work by Costa $\sl{et. al.}$~\cite{PhysRevD.85.024001}, where it is shown that the mass center of a spinning particle is observer-dependent in relativistic physics. For two different observers, with the relative velocity $\bm v$ between them, the mass center of a spinning object has a shift
\be
\delta\bm x=\frac{\bm S_{\star}\times\bm v}{M},\non
\ee
and this is similar to the shift in Eq.~(\ref{eq10}). Closely related results can also be found in the work by Gralla, Harte and Wald~\cite{PhysRevD.81.104012} and in the textbook by Misner, Thorn, and Wheeler~\cite{misner1973gravitation}.

In the work by Chang and Niu~\cite{RevModPhys.82.1959}, with the non-relativistic limit and $\bm B=0$, the velocity of a Dirac wave packet is estimated to be
\be
\frac{d\bm r_\mathrm{c}}{dt}=\bm v-\frac{e}{2m^2}\bm s\times \bm E,\non
\ee
and the absolute  value of  the anomalous velocity above equals  $\vert\bm V_{\mathrm{P}}(\bm E)\vert$ derived in this work. We would like to emphasize that the Berry's curvature, which was previously introduced by Chang and Niu~\cite{0953-8984-20-19-193202}, does not appear in our expression of the anomalous velocity. Moreover, from the work by Mathur~\cite{PhysRevLett.67.3325}, one can know that Berry's curvature has a deep relationship with Thomas precession. As shown in Eq.~(\ref{eq11}), it is obvious that the Thomas precession has a contribution to the anomalous velocity. Thus, how the anomalous velocity is related to Berry's curvature could be discussed in the future work.

Finally, attention may be paid to the definitions of the mass center and rotational kinematics for relativistic $N$-body problem, which goes beyond the results given by Pryce. A more detailed discussion about this topic can be found in the work by Alba, Lusanna  and
 Pauri~\cite{10.1063/1.1416889}.  The Pryce's e-type operator (the Newton-Wigner position operator or the mean-position operator), plays a fundamental role in quantum mechanics, and this is one of the most important reasons for us to focus on the Pryce's mass-center operators. It was emphasized by Foldy and Wouthuysen that this operator corresponds to the position of a particle in Pauli theory, and in our present work, we emphasize that the mean-position operator predicts a non-vanishing anomalous velocity. However, it is still an open question that how Pryce's mass-center operators are related to the anomalous velocity of the Bloch electron .

Let us end our work with a short conclusion. Based on the argument that the mass center of a classical spinning electron should be associated with the expectation value of the Pryce's mass-center operator, within the scheme developed  by Fradkin and Good, the anomalous velocity of the spinning electron is shown to be determinated by the position shift and the selection of the mass-center operators.

\section*{acknowledgements}
The authors would like to thank Prof. Xiaofu L\"u, Prof. Yan He, and Dr. Xiaozhao Chen for the helpful discussions. This work was supported by the National Natural Science Foundation of China under Grant No.~11405136, and the Fundamental Research Funds for the Central Universities under Grant No.~2682016CX059.

\bibliography{refs}

\end{document}